\documentclass[reprint]{JASA}
\usepackage{tipa}
\usepackage{setspace}
\usepackage{array}
\usepackage{hyperref}
\usepackage{soul}
\usepackage{multirow}
\usepackage{textcomp}
\usepackage{enumitem}
\usepackage{ulem}
\newcommand{\mytilde}{\raise.17ex\hbox{$\scriptstyle\mathtt{\sim}$}}
\newcolumntype{L}[1]{>{\raggedright\let\newline\\\arraybackslash\hspace{0pt}}p{#1}}
\newcolumntype{C}[1]{>{\centering\let\newline\\\arraybackslash\hspace{0pt}}p{#1}}
\newcolumntype{R}[1]{>{\raggedleft\let\newline\\\arraybackslash\hspace{0pt}}p{#1}}
\definecolor{pureyellow}{rgb}{1, 1, 0}
\definecolor{cyanblue}{rgb}{0, 1, 1}
\definecolor{revred}{rgb}{0.86, 0.35, 0.35}
\newcommand{\hlbl}[1]{\sethlcolor{cyanblue}\hl{#1}}
\newcommand{\hlye}[1]{\sethlcolor{pureyellow}\hl{#1}}

\begin{document}

\title[]{Lexical and syntactic gemination in Italian consonants - Does a geminate Italian consonant consist of a repeated or a strengthened consonant?}

\author{M.-G. Di Benedetto}
\affiliation{Massachusetts Institute of Technology (MIT), Cambridge, MA, United States,\\ DIET Department,  Sapienza University of Rome, Rome, Italy}
\author{Stefanie Shattuck-Hufnagel}
\affiliation{Massachusetts Institute of Technology (MIT), Cambridge, MA, United States}

\author{Luca De Nardis}
\author{Sara Budoni}
\affiliation{DIET Department,  Sapienza University of Rome, Rome, Italy}

\author{Javier Arango}
\author{Ian Chan}
\author{Alec DeCaprio}			
\affiliation{Radcliffe Institute for Advanced Study, Harvard University, Cambridge, MA, United States}

\preprint{Di Benedetto et al., JASA}		

\date{\today} 




\begin{abstract}
Two types of consonant gemination characterize Italian: lexical and syntactic. Italian lexical gemination is contrastive, so that two words may differ by only one geminated consonant. In contrast, syntactic gemination occurs across word boundaries, and affects the initial consonant of a word in specific contexts, such as the presence of a monosyllabic morpheme before the word. This study investigates the acoustic correlates of Italian lexical and syntactic gemination, asking if the correlates  for the two types are similar in the case of stop consonants. Results confirmed previous studies showing that duration is a prominent gemination cue, with a lengthened consonant closure and a shortened pre-consonant vowel for both types. Results also revealed the presence, in about 10-12\% of instances, of a double stop-release burst, providing strong support for the biphonematic nature of Italian geminated stop consonants. Moreover, the timing of these bursts suggests a different planning process for lexical vs. syntactic geminates. The second burst, when present, is accommodated within the closure interval in syntactic geminates, while lexical geminates are lengthened by the extra burst. This suggests that syntactic gemination occurs during a post-lexical phase of production planning, after timing has already been established.
\end{abstract}


\maketitle



\section{Introduction}
Consonant gemination is the process by which a consonant is produced, as the word “gemination” hints, as “doubled”, that is, as two consecutive occurrences of the same phoneme, or, under a different interpretation, as a stronger, longer, or more intense, consonant. 
Geminate consonants are present in several languages such as Italian \cite{EspDiB99}, Japanese \cite{HirWhi05}, Arabic \cite{AlT15}, Russian \cite{Dmi17}, and Persian \cite{Han16}. In some of these languages, and in particular in Italian, gemination is contrastive, that is, the lexicon of these languages includes minimal word pairs in which the meaning of one word is distinguished from its minimal pair counterpart on the sole basis of consonant gemination. In Italian this contrast is very widely used and numerous minimal pairs are present in the lexicon, as, for instance, \emph{pala} vs. \emph{palla} (shovel vs. ball) or \emph{pena} vs. \emph{penna} (pain vs. pen).\\

In Italian -- see the examples above -- when a geminate consonant appears within a word, it is usually orthographically transcribed as two consecutive graphemes of the same consonant. This is the case in Italian for most consonants: stop consonants as well as a subset of nasals and fricatives. As a matter of fact, most Italian consonants can be geminated in intervocalic position, with the exception of a few such as /z/, although different experts of Italian phonology hold contrasting views regarding a  particular subset of five consonants /ts, \textdzlig, \textesh, \textltailn, \textlambda /  (\cite{Por39} vs. \cite{Mul72}). Throughout this study, we will characterize gemination properties in agreement with what is proposed by Muljacic (1972), and, in particular, we will assume that all Italian consonants except /z/ can be geminated, although the above five consonants do have a special status, in that these particular consonants are always geminated in intervocalic position and there exist no minimal pairs based on the contrastive gemination effect. For these five consonants, the orthographic transcription makes use of the presence of either one or two graphemes, as in the words \emph{azione} (/ats’tsjone/) (action) vs. \emph{polizza} (/’politstsa/) (policy), for instance, although /ts/ is acoustically geminated in both words. Table \ref{tab:phonemes} shows a list of the Italian consonants and of their geminate counterparts, and summarizes the specific properties of the different consonants.  

\begin{table*}[t]
\caption{\label{tab:phonemes}List of Italian consonants and their gemination behavior. For each consonant an example of a word containing it, the IPA phonemic and geminate transcriptions, and typical properties of occurrence are given.}
\begin{ruledtabular}
\centering\footnotesize\setstretch{0.7}
\begin{tabular}{C{1.5cm}C{1.5cm}C{2cm}C{2.5cm}C{6.5cm}}
Grapheme	& Example of Word & IPA Phonemic Transcription & IPA Phonemic Geminate Transcription & Occurrence\\
\hline
n & nonna &	/n/ & /nn/ & single and geminated form intervocalically\\
r & ragazzi & /r/	& /rr/	& single and geminated form intervocalically\\
t & teoria & /t/	& /tt/ &	single and geminated form intervocalically\\
d & digitale & /d/& /dd/	& single and geminated form intervocalically\\
l & lavoro & /l/ & /ll/ & single and geminated form intervocalically\\
s & sorelle & 	/s/ & 	/ss/ & single and geminated form intervocalically\\
c & cugino & /k/	 & /kk/ & single and geminated form intervocalically\\
p & parole & 	/p/ & /pp/ & single and geminated form intervocalically\\
m & mattino	 & /m/ & /mm/ & single and geminated form intervocalically\\
v & vacanza & /v/ & /vv/ & single and geminated form intervocalically\\
ci, ce & città & /\textteshlig /& /\textteshlig \textteshlig/ & single and geminated form intervocalically\\
f & fiamme & /f/	 & /ff/ & single and geminated form intervocalically\\
g & gatto & /g/ & /gg/ & single and geminated form intervocalically\\
b & bambino & /b/ & /bb/ & single and geminated form intervocalically\\
gi & giardino & /\textdyoghlig/ & /\textdyoghlig \textdyoghlig/ & single and geminated form intervocalically\\
z & zitto & /\texttslig/ & 	/\texttslig \texttslig/ & 	only in geminated form intervocalically\\
gl & figlio & /\textlambda/	 & /\textlambda \textlambda / & only in geminated form intervocalically\\
sci & scienzato & /\textesh/ & /\textesh \textesh/ & only in geminated form intervocalically\\
z & zoo & /\textdzlig/ & /\textdzlig \textdzlig/ & only in geminated form intervocalically\\
s & svetta & /z/ & N/A & never in geminated form\\
gn	 & gnomi & /\textltailn/ & /\textltailn \textltailn/ & only in geminated form intervocalically\\
\end{tabular}
\end{ruledtabular}

\end{table*}

Although gemination is found in many languages, in Italian it has a peculiar property that distinguishes it from many others. In Italian, in fact, gemination may reflect a kind of assimilation across word boundaries in particular circumstances, giving rise to the so-called syntactic gemination effect (in Italian \emph{Raddoppiamento Sintattico}, RS). This phenomenon is widely used in Italian compared to the very few other languages that show a similar effect, such as Finnish and in some way Maltese for Italian and Sicilian imported words.  In RS, the initial consonant of a word, that in standard Italian is always a single consonant, becomes geminated when that word is preceded by a monosyllabic morpheme, for example a function word, or if the preceding word has its lexical accent on the last syllable. For example, in the group of words \emph{a piedi} (by foot), the initial consonant of piedi /p/ becomes geminated, so that the phonemic transcription of the post-lexical word group is /ap’pj\textepsilon di/. Although it is not within the scope of this paper to describe in detail all the specific cases in which syntactic gemination may take place in Italian (for a comprehensive analysis see \cite{Cam65}, pp. 133-154) -- but rather to introduce the phenomenon in order to include it in the study -- it is interesting to note that syntactic gemination in Italian can be contrastive with respect to pairs of word groups; An interesting example is the case of the group of words \emph{tra monti} (among mountains), in which /m/ is geminated and the post-lexical word group is transcribed as /tram’monti/, vs. the word \emph{tramonti} (sunsets) that is transcribed as /tra’monti/. The gemination of the /m/ consonant distinguishes a post-lexical word group from a word of the lexicon. \\
To sum up, Italian is characterized by two types of gemination: lexical and syntactic. In standard Italian, lexical geminate consonants only occur within words (that is, never in initial position), while initial consonants of words may become geminated due to the syntactic gemination effect. It should be noted that in some dialects of southern and central Italy consonants may be also geminated in word initial position, independently of syntactic gemination (\cite{BerLop05}; \cite{Bon11}), but in this case the effect is not contrastive and seems to be more of a phenomenon involving junction and adjustment between consonants occurring at word boundaries. In this instantiation, gemination seems to resemble the typical phenomenon of \emph{liaison} in French, which is known to occur more often with words that co-occur frequently but does not prove, when realized, to favor lexical recognition of linked words \cite{FouGol01}.  This result leads to an interpretation by which non-contrastive gemination, like \emph{liaison}, may be considered as an expressive sandhi phenomenon.\\

As mentioned above, consonant gemination is usually described either as the doubling of a consonant, or as the production of a single consonant that is stronger or more intense and typically characterized by longer duration. Whether a geminate consonant is represented in the mind of the speaker as a single longer or stronger consonant vs. a double consonant has been debated for several decades \cite{Swa37}, and, at least for the Italian language, is still under discussion. These two opposite views, by which a geminate consonant is interpreted as either one /C:/ or two /CC/ phonemes, may lead to two different ways of considering the syllabic structure of /CC/, either heterosyllabic /C.C/ or tautosyllabic  /C:/. These two different views have possible relevant consequences on the extent of coarticulation effects between syllables \footnote{It is interesting to note that Latin, from which Italian derives and to which is the closest one among the Romance languages, had contrastive lexical consonant gemination, as well as expressive consonant gemination, and that there is an almost unanimous consensus that in Latin geminate consonants were actually two consonants \cite{GiaMar89}.}. 

The search for acoustic correlates of gemination in Italian, and the verification of their perceptual relevance, has been the object of a longstanding project, the Gemination project GEMMA \cite{DiB00}, \cite{GEM19}.  This project began at Sapienza University of Rome in 1992, with the goal of analyzing gemination in Italian consonants, based on the analysis of VCV vs. VCCV words. Results for stops \cite{EspDiB99}, fricatives and affricates \cite{DiBDeN20b}, and nasals and liquids \cite{DiBDeN20a}, showed a general tendency to shorten the pre-consonant vowel and to lengthen the word-medial consonant in a geminate word. No significant effects of gemination were observed on other acoustic parameters, such as energy- and frequency-related measurements. The general conclusion was that consonant duration is a primary cue to gemination, and pre-consonant vowel duration a secondary cue. \\

This article addresses the problem of characterizing lexical and syntactic gemination in Italian in terms of acoustic manifestations and acoustic correlates, and of understanding whether these correlates are similar between lexical and syntactic geminates. The analysis was carried out on spoken sentences in which both lexical and syntactic gemination occurred. The questions were: a) Do geminates in running speech manifest, as observed in previous studies, mostly by varying time-based parameters? And is it possible to disentangle the question about the biphonematic vs. monophonematic nature of Italian geminate consonants, by observing the acoustic manifestation of gemination in running speech? b) Do lexical and syntactic geminates organize the temporal distribution among segments in a similar way? c) How do the findings impact cluster syllabification? In other words, does gemination sometimes result in a longer single consonant but in other cases in a doubled consonant? And if so, when the geminated consonant consists of two consecutive consonants $\mathrm{C^{(1)}C^{(2)}}$ rather than one stronger and longer consonant, do $\mathrm{C^{(1)}}$ and $\mathrm{C^{(2)}}$ differ acoustically, and is $\mathrm{C^{(2)}}$ stronger and more stable than $\mathrm{C^{(1)}}$? A positive answer may lead to the hypothesis that the sequence $\mathrm{C^{(1)}C^{(2)}}$ is heterosyllabic, $\mathrm{C^{(1)}}$ being a coda consonant and $\mathrm{C^{(2)}}$ an onset consonant. The answer to this specific question may lead to improved understanding of the production planning process.\\

The paper is organized as follows. Section \ref{sec:experimentation} contains the description of the database and of the experiment. Section \ref{sec:analysis} contains the results of the acoustic analysis. Section \ref{sec:discussion} contains a general discussion of results and the proposed interpretation, as well as the conclusions. 

\section{Experimentation: gemination in spoken sentences}
\label{sec:experimentation}

\begin{figure*}[ht]
\includegraphics[width=\textwidth]{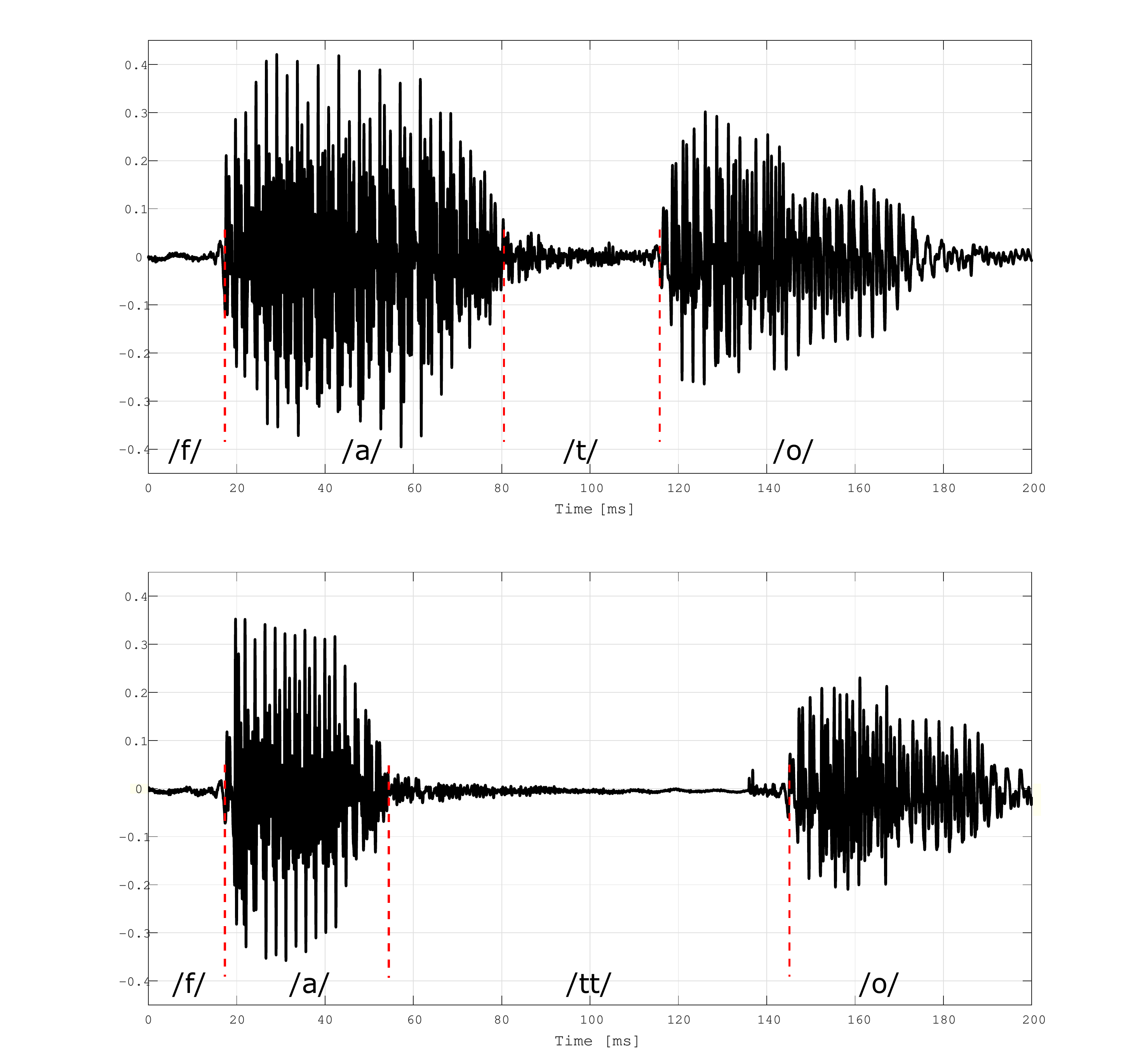}
\caption{\label{fig:FIG1}{Singleton stop /t/ vs. geminated stop /tt/ in words \emph{fato} vs. \emph{fatto} as pronounced in running speech in sentences \emph{Il fato ancora} vs. \emph{Il fatto ancora}, highlighting the lengthening of consonant and shortening of preceding vowel associated to gemination, as described in \cite{EspDiB99}. }}
\raggedright
\end{figure*}

As mentioned in the Introduction, previous studies of consonant gemination in Italian VCV and VCCV words, showed that the contrast between singleton vs. geminate consonants was durational in nature. In particular, these studies indicated that the duration of the consonant and the duration of the pre-consonant vowel are the two parameters that are significantly different for the two consonant categories. An example of the impact of gemination on the parameters is presented in Figure \ref{fig:FIG1}, which shows the waveforms associated with the words \emph{fato} vs. \emph{fatto} as pronounced in running speech in sentences \emph{Il fato ancora} vs. \emph{Il fatto ancora}.

Based on these previous studies, acoustic analysis carried out in this experiment aimed at measuring these specific parameters for both lexical and syntactic geminates.\\
In the experiment design, stop consonants were selected as the consonants to be measured, following an approach that was also adopted in the GEMMA project, since stops are the most frequent geminated consonants in Italian and are also the most informative; Not only are these consonants easier to measure with fine detail, thanks to a clear presence of closure and release phases, but also, in stops there is a possibility that if the biphonematic hypothesis were true one could eventually find evidence of two bursts, two closures, and two releases.
The speech material on which the analysis was carried out consisted of 100 Italian spoken sentences forming the LaMIT database \cite{DiBCho20}. The set of sentences is reported in Table \ref{tab:LaMIT_sentences}. 

\begin{table*}[t]
\caption{\label{tab:LaMIT_sentences}Sentences of the LaMIT database.}
\begin{ruledtabular}
\centering\setstretch{0.5}
\scriptsize
\begin{tabular}{L{8.3cm}|L{8.3cm}}
1. Il gatto della vicina è bianco peloso e pazzo & 51. Sono belli i programmi decisi all’ultimo momento\\
2. Il giardino di mio cugino è pieno di gladioli e di gnomi & 52. Con Cristiana pratico yoga ogni mercoledì\\
3. L’università italiana è un’istituzione pubblica dello stato & 53. Pensieri e parole cantava la diva con voce suadente\\
4. Passeggerei volentieri a piedi nudi nella città vecchia  & 54. Aprile si esaurisce mentre arriva carico di promesse il mese di maggio\\
5. Pietro non scappa fugge a gambe levate con il cuore in fiamme & 55. Abbiamo trasmesso il giornale radio del mattino\\
6. Cosa ne penseresti di alzarti presto e salutare il sole & 56. Il tempo previsto sull’Italia per questa sera non prevede temperature in aumento\\
7. Quando Maria è in vacanza compra volentieri la settimana enigmistica & 57. Pensavo che tu volessi fare solo uno spuntino\\
8. Lo schermo del tuo cellulare è graffiato e opacizzato & 58. Assicurati che non si dimentichino di scrivere alla zia\\
9. All’imbrunire la cattedrale svetta nel cielo basso e uggioso & 59. Per salvarci dobbiamo restare uniti\\
10. Alcuni studenti dell’anno accademico corrente potranno laurearsi a luglio & 60. Il mondo è nelle nostre mani \\
11. Due sorelle si aiutano se vanno d’amore e d’accordo & 61. Comportati educatamente a tavola\\
12. La struttura precaria resse malgrado il forte vento & 62. Pare che sia rimasto solo per un colpo di testa\\
13. Mandare cartoline da città remote non è più di moda & 63. La piccola peste vuole il ciuccio per calmarsi\\
14. Discendi il Monte Bianco con gli sci e vivi un’esperienza unica e indimenticabile & 64. Non mordere la spalla della nonna\\
15. Prima o poi dovrai pur deciderti a leggere le opere di Niccolò Machiavelli & 65. La carta non si mangia se non sei una capra\\
16. Non potendo fare a meno del cioccolato pensò bene di privarsi della panna montata & 66. Il pavone becca le foglie sul viale dello zoo\\
17. Che avventura meravigliosa quella di guardare gattonare un bebè & 67. All’improvviso si udì l’urlo del barbagianni\\
18. “E pur si muove” disse il famoso scienziato rivolgendosi agli inquisitori & 68. Basta con i fanatismi esagerati\\
19. Oggi piove a dirotto governo ladro & 69. Non smettere di fantasticare ad occhi aperti\\
20. Riporre tanti sogni nel cassetto rinforza la fantasia del poeta & 70. Col vento in poppa attraversarono il Mediterraneo in un soffio\\
21. Vent’anni di allenamento non furono sufficienti a chiudere la pinza & 71. Voltati e renditi conto di quanta strada hai percorso\\
22. Senti un po’ di musica e vedi che ti passa la nostalgia dell’inverno & 72. Una tazza di te verde al giorno rinfresca la mente\\
23. I clienti della Banca devono attenersi alle regole stabilite dal contratto & 73. Scriverò questa lettera con la penna a sfera\\
24. Apponi la firma in calce perché è necessario per rendere valida la transazione & 74. Che ne pensi di una fetta di torta\\
25. I ragazzi della scuola religiosa fisseranno un appuntamento con il sindaco ateo & 75. Il cestino per la carta sta sotto la scrivania\\
26. Se prendi in prestito un libro alla biblioteca godi del vantaggio di non dover acquistarlo & 76. Una vacanza in agriturismo in Toscana ha un costo ridotto\\
27. La rappresentazione digitale delle immagini ha rivoluzionato la fotografia & 77. Sul pavimento del salone giace un tappeto persiano\\
28. Addio all'imperatore giapponese abdicherà oggi in favore di suo figlio & 78. L’albero di cedro è simbolo del Libano\\
29. Uno sciame di api investì il bambino biondo costringendolo a buttarsi giù dall’albero & 79. Un biglietto di auguri accompagna il regalo\\
30. La ferrovia si snoda lungo il fiume seguendo un tracciato tortuoso & 80. Torneresti a casa a piedi\\
31. Dopo avere letto molti libri Luca si rimise a studiare ancora per un po’ & 81. Il grano saraceno non contiene glutine\\
32. Se arrivi all’alba a Capri butta l’ancora e prosegui a nuoto & 82. Il pane lievita quando la luna è piena\\
33. Giorgio ha deciso di prendere i voti ma prima ha dovuto battezzarsi & 83. Stendi il bucato al sole e risparmi energia\\
34. Che ne farai dei quaderni di storia & 84. Sotto la piazza giace un tesoro\\
35. Chiedi pure a tuo padre cosa ne pensa dell'anguria & 85. La balena blu nuota in solitario\\
36. Mamma e papà ti vogliono bene & 86. Servono nuovi dirigenti per rilanciare le aziende\\
37. Non poggiare il bicchiere colmo d’acqua sul pianoforte  & 87. Creare lavoro è un dovere costituzionale\\
38. Con la bicicletta elettrica le salite sono una passeggiata & 88. Ma questa è un’altra storia su cui si indagherà\\
39. Saluta la signora e fai l’inchino & 89. Si al regolamento che impone limiti alla stupidità\\
40. La teoria dei numeri è una branca della matematica & 90. La ballerina indossa un costume rosa fragola\\
41. Mio nipote ama trovare soluzioni a problemi complessi & 91. L’autore si muove con scioltezza nella palude delle parole\\
42. Aguzza l’ingegno e progetta una radio intelligente & 92. La fascetta giusta dovrebbe essere alienazione\\
43. Il giornalaio vende e invia riviste e oggetti turistici & 93. Resti in collegamento che risponderà il primo operatore libero\\
44. Il cane corse forsennatamente verso il padrone calpestando le aiuole & 94. Vediamo se la risposta è quella giusta\\
45. Il dolore sorgeva mentre la luna non era ancora tramontata & 95. L’avocado cresce nei paesi tropicali\\
46. Puoi accendere la radio a caso e sintonizzarti su qualsiasi frequenza & 96. Pesce fritto e insalata mista grazie\\
47. Poi ci sono i rimedi naturali che sono più efficaci di tanti prodotti presenti in farmacia & 97. Favorisce un caffè dopo cena col digestivo\\
48. Impariamo a meditare giornalmente & 98. La folla era impazzita alla vista dell’assassino\\
49. Si perde così tanto tempo a discutere del niente & 99. Lei col maglione rosso si stia zitto\\
50. Stasera andremo al cinema a vedere un film francese & 100. Mangerebbe volentieri un filetto di baccalà con le olive\\
\end{tabular}
\end{ruledtabular}
\end{table*}

The sentences were designed to include all the phonemes of the Italian language and the geminate versions of the consonants (except for /z/, since this consonant is not geminated, as mentioned in the Introduction). Take for instance the first sentence of Table \ref{tab:LaMIT_sentences}, in which both lexical and syntactic geminates occur. By highlighting lexical geminates in cyan and syntactic geminates in yellow, the first sentence \emph{Il gatto della vicina è bianco peloso e pazzo} is transcribed as follows:\\

/ {\textquotesingle}il {\textquotesingle}ga\hlbl{tt}o {\textquotesingle}de\hlbl{ll}a vi{\textquotesingle}\textteshlig ina {\textquotesingle}\textepsilon \hphantom{} \hlye{b{\textquotesingle}b}janko pe{\textquotesingle}loso {\textquotesingle}e \hl{p{\textquotesingle}p}a\hlbl{\texttslig \texttslig}o / \\

The LaMIT database was designed to reflect the typical frequency of occurrence of the different phonemes in the Italian language, as suggested by a recent study \cite{AraDeC21}, \cite{AraYao20} that provides updated values of the phonemic frequencies, assuming the existence of 30 phonemes as identified by Muljacic (1972): 1) seven vowels: /a/, /i/, /u/, /e/, /\textepsilon/, /o/, /\textopeno/; 2) twenty-one consonants: /p/, /b/, /f/, /v/, /t/, /d/, /\texttslig/, /\textdzlig/, /s/, /z/, /k/, /g/, /\textdyoghlig/, /\textteshlig/, /\textesh/, /m/, /n/, /\textltailn/, /l/, /\textlambda/, /r/; 3) two glides: /j/ and /w/. Allophones are excluded, consistent with the theoretical framework provided by Muljacic (1972).\\

Speech materials were recorded in the Speech Laboratory of the DIET Department at Sapienza University of Rome, Rome, Italy, on a MacBook Pro laptop connected to a Samson Meteor Mic USB microphone, using the Audacity software tool with a sampling rate of 44.1 kHz and quantization set at 16 bits per sample. All recordings were performed in a sound-treated room under the supervision of an acoustically trained person. The speakers were Italian native speakers, raised and living in Rome (Italy), pronunciation defectless with no reported speech disorder and free of evident dialectal inflections. The supervisor was an Italian native speaker as well, from Naples but living in Rome for many years. As suggested in \cite{Pay06}, the Roman accent, although quite distinctive, is phonologically very close to Standard Italian. The entire set of sentences was recorded twice in two different recording sessions, leading to two repetitions for each sentence per speaker. In case of evident mispronunciations, as for example a missed or wrong word, the speaker was asked to repeat the sentence.\\
The speech materials recorded by one male speaker (MS, age 45) and one female speaker (FS, age 25), formed the object of the present analysis. 

\section{Acoustic analysis}
\label{sec:analysis}

\begin{figure*}[t]
\baselineskip=12pt
\figcolumn{
\fig{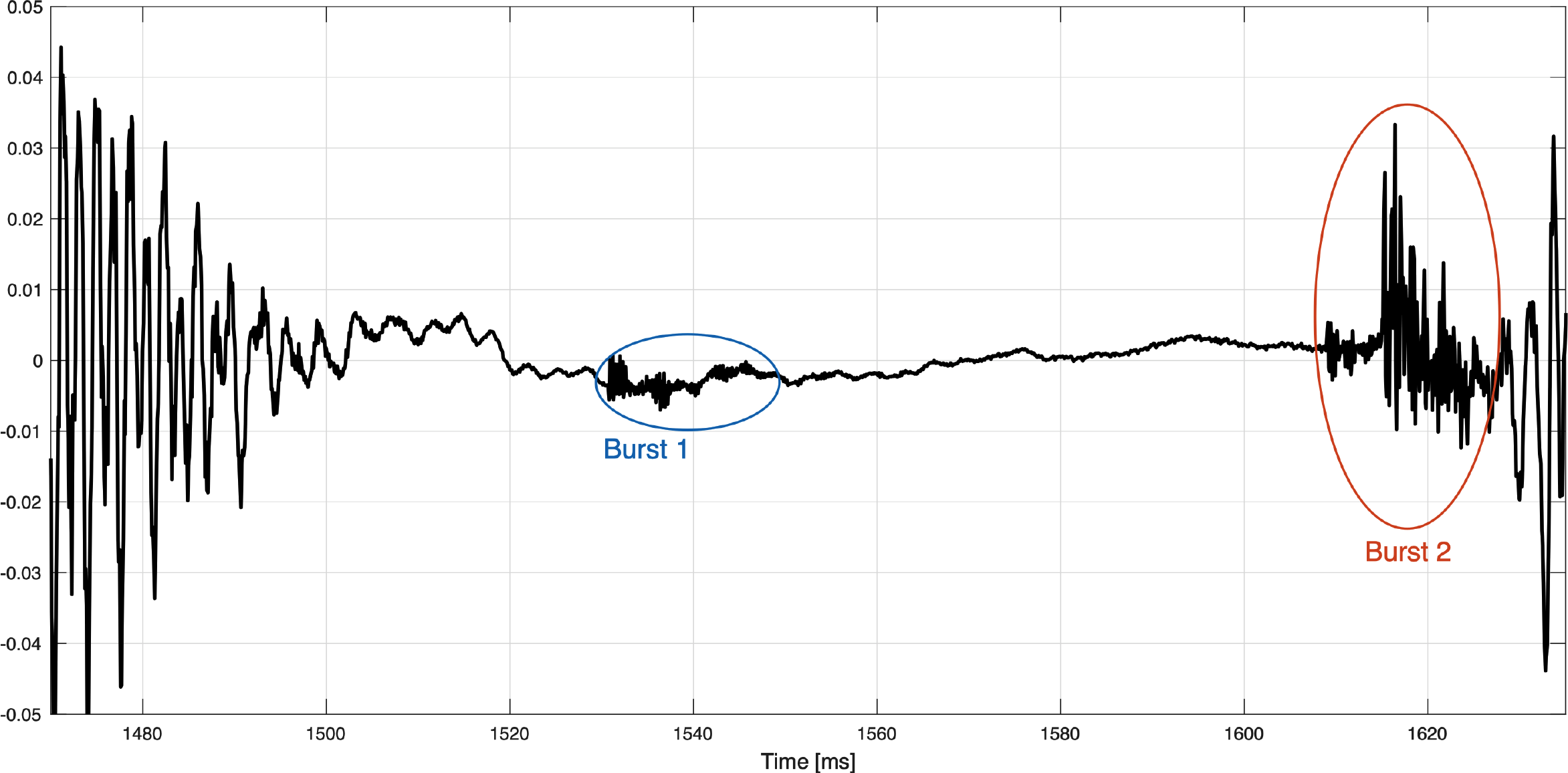}{.9\textwidth}{(a)}\label{fig:Fig2a}
\fig{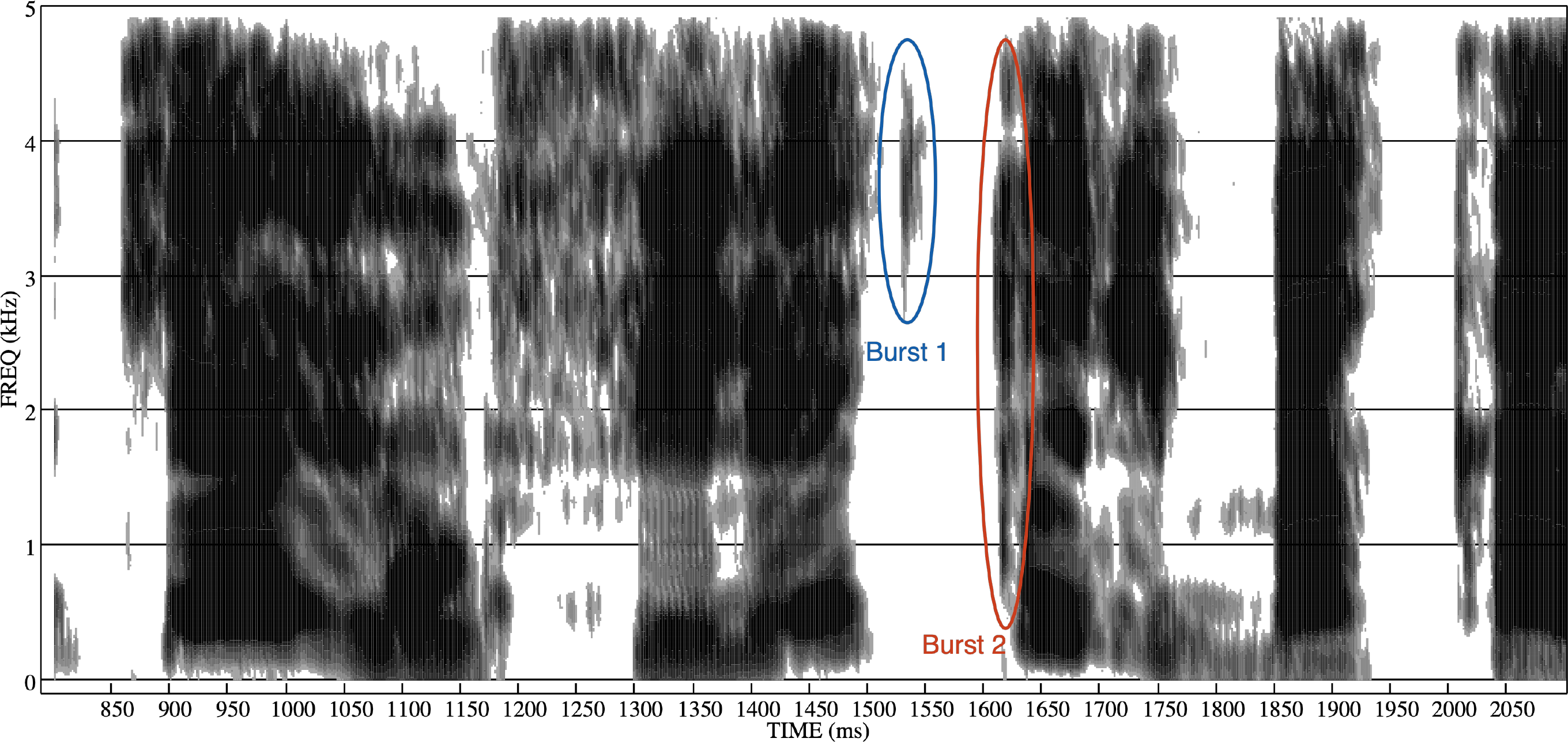}{.9\textwidth}{(b)}\label{fig:Fig2b}
}
\caption{\label{fig:Fig2}Example of double burst: the geminate [t] in the word \emph{filetto} of sentence 100 of speaker MS, repetition 1. Waveform (Fig. \ref{fig:Fig2a}) and spectrogram (Fig. \ref{fig:Fig2b}) show a first burst at time {\mytilde}1530 ms to {\mytilde}1549 ms and a second burst at time {\mytilde}1610 ms to {\mytilde}1632 ms. Note that waveform is zoomed on the consonant portion to make bursts more visible. }
\end{figure*}

The acoustic analysis was conducted manually by examining the signal and the spectrogram of all sentences, for both repetitions and speakers.   Speech signals were analyzed using the xkl software, part of the set of software tools developed by Dennis Klatt \cite{Kla84}. All sentences were also checked by listening to confirm the presence of gemination.\\
A substantial number of instances of double closures and bursts were observed for both speakers MS and FS, providing evidence for the presence of two consecutive consonants $\mathrm{C^{(1)}}$ and $\mathrm{C^{(2)}}$. Double bursts were found in about 12\% of instances of lexical geminates and 10\% of syntactic geminates. Table \ref{tab:doublegem} shows the distribution of single vs. double bursts for lexical vs. syntactic gemination in both speakers. Although a similar number of double bursts was found for the two speakers, in both syntactic and lexical forms, the observed double bursts occurred in different sentences and for different consonants. 
\begin{table*}[t]
\caption{\label{tab:doublegem}Number of single burst vs. double burst geminates in both lexical and syntactic forms and for each speaker.}
\begin{ruledtabular}
\centering\footnotesize\setstretch{0.7}
\begin{tabular}{C{3cm}|C{2cm}|C{2cm}|C{1.5cm}|C{2cm}|C{2cm}|C{1.5cm}}
 & \multicolumn{3}{c|}{Lexical gemination} & \multicolumn{3}{c}{Syntactic gemination}\\
 \hline
 & single burst & double burst & total	& single burst & double burst	& total\\
speaker MS	&105 & 15 & 120 & 69 & 7 & 76\\
speaker FS & 105 & 15 & 120 & 68 & 8 & 76\\
Total	& 210 & 30 & 240 & 137 & 15 & 152\\
\end{tabular}
\end{ruledtabular}
\end{table*}
Table \ref{tab:doublegemdetail} shows where a double burst was present -- in which sentences and which words. A typical example of an observed double burst is presented in Fig. \ref{fig:Fig2}, showing the waveform and spectrogram of the geminate [t] in the word \emph{filetto} of sentence 100 of speaker MS, repetition 1. As shown in Fig. \ref{fig:Fig2}, a first burst appears at time {\mytilde}1530 ms to {\mytilde}1549 ms. A second burst is visible at time {\mytilde}1610 ms to {\mytilde}1632 ms.
\begin{table*}[t]
\caption{\label{tab:doublegemdetail}Inventory of words and word groups, and corresponding LaMIT database sentence number and  repetition, in which double burst lexical and syntactic geminates were found.}
\begin{ruledtabular}
\centering\footnotesize\setstretch{0.7}
\begin{tabular}{C{3cm}|C{1.5cm}|C{2.5cm}|C{3cm}|C{1.5cm}|C{2.5cm}}
 \multicolumn{6}{c}{Lexical gemination} \\
  \hline
  \multicolumn{3}{c}{Speaker FS} & \multicolumn{3}{c}{Speaker MS} \\
 \hline
 sentence number & repetition & word & sentence number & repetition & word\\
1 & 1 & ga\textbf{tt}o & 4 & 1 & ve\textbf{cc}hia\\
9 & 1 & 	sve\textbf{tt}a & 12 & 1 & stru\textbf{tt}ura\\
13	 & 1 & ci\textbf{tt}à & 15	 & 1 & Ni\textbf{cc}olò\\
15 & 	1 & Ni\textbf{cc}olò & 23 & 1 & contra\textbf{tt}o\\
20 & 1 & casse\textbf{tt}o & 38 & 1 & ele\textbf{tt}rica\\
63 & 1 & pi\textbf{cc}ola & 63 & 1 & pi\textbf{cc}ola\\
100 & 1 & file\textbf{tt}o & 69 & 1 & sme\textbf{tt}ere\\
3 & 2 & pu\textbf{bb}lica & 69 & 1 & o\textbf{cc}hi\\
4 & 2 & ci\textbf{tt}à & 100 & 1 & file\textbf{tt}o\\
5 & 2 & sca\textbf{pp}a & 4 & 2 & ve\textbf{cc}hia\\
7 & 2	 & se\textbf{tt}imana & 9 & 2 & sve\textbf{tt}a\\
19	 & 2 & diro\textbf{tt}o & 15	 & 2 & Ni\textbf{cc}olò\\
69 & 2 & sme\textbf{tt}ere & 19 & 2 & diro\textbf{tt}o\\
92 & 2 & fasce\textbf{tt}a & 32 & 2 & bu\textbf{tt}a\\
99 & 2 & zi\textbf{tt}o & 27 & 2 & bi\textbf{cc}hiere\\
 \hline
 \multicolumn{6}{c}{Syntactic gemination} \\
  \hline
  \multicolumn{3}{c}{Speaker FS} & \multicolumn{3}{c}{Speaker MS} \\
 \hline
 sentence number & repetition & word & sentence number & repetition & word\\
19 & 1 & a \textbf{d}irotto & 5 & 1 & a \textbf{g}ambe\\
21 & 	1 & a \textbf{c}hiudere & 8 & 1 & è \textbf{g}raffiato\\
33 & 1 & ha \textbf{d}eciso & 21 & 1 & a \textbf{c}hiudere\\
35	 & 1 & a \textbf{t}uo & 33 & 1 & ha \textbf{d}eciso\\
80 & 1 & a \textbf{c}asa & 94 & 1 & è \textbf{q}uella\\
21 & 2 & a \textbf{c}hiudere & 33 & 2 & ha \textbf{d}eciso\\
22	 & 2 & po’ \textbf{d}i & 94 & 2 & è \textbf{q}uella\\
32	 & 2 & a \textbf{c}apri & - & - & 	-\\
\end{tabular}
\end{ruledtabular}
\end{table*}

\begin{figure*}[t]
\includegraphics[width=\textwidth]{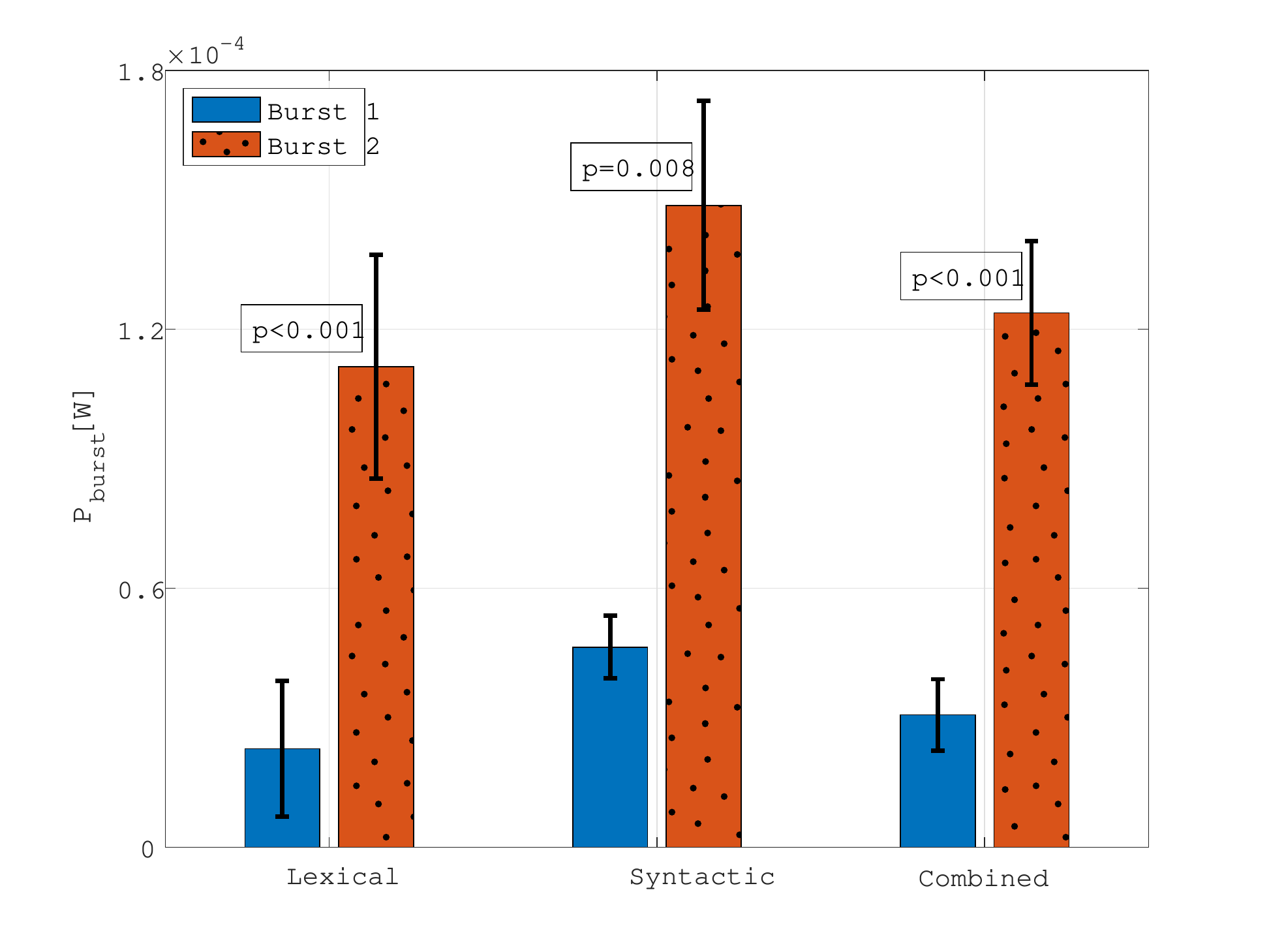}
\caption{\label{fig:Fig3}Average value and standard error of power of the first burst (blue, plain) vs. the second burst (red, dotted), for lexical and syntactic groups, and for both groups combined. The power of the burst was computed as the energy of the burst divided by the number of samples composing it. Values of $p$ in bold indicate that $p<p*=0.05$, i.e., a statistical significance of the parameter. }
\raggedright
\end{figure*}

For the instances outlined in Table~\ref{tab:doublegem}, which we respectively call lexical and syntactic double bursts, a comparative analysis showed that the second burst was stronger than the first, as also visible on Fig. \ref{fig:Fig2}. To quantify this observation, the power of the burst was computed as the energy of the burst divided by the number of samples composing it, that is:
\begin{equation}
    P_{burst}=\frac{1}{N}\sum_{i=1}^{N}x_i^2,
    \label{eq:p_burst}
\end{equation}
where $x_i$ is the $i$-th speech sample amplitude and $N$ is the number of speech samples in the burst. Three ANOVA univariate tests on the $P_{burst}$ parameter, where the fixed factor was first burst vs. second burst were then carried out: one test for each gemination form, i.e. lexical vs. syntactic, and one test with lexical and syntactic forms combined. Given the number of available samples (30 lexical cases and 15 syntactic cases) the threshold for significance was set at $p*=0.05$. Results are presented on Fig. \ref{fig:Fig3}, showing that the second burst was significantly stronger than the first, for each form separately, and also combined. Table~\ref{tab:ANOVA_power} in the Appendix contains the details of the statistical tests; in particular, F values, degrees of freedom, p values, and the effect size parameter $\eta^2$. Parameter $\eta^2$ complements the information provided by p; when p is below the significance threshold, $\eta^2$ provides an indication of the extent of the difference between the two groups separated by the factor. One possible criterion for interpreting $\eta^2$ was proposed by \cite{Coh88}, and classifies the effect as small for $0.0099<\eta^2<0.0588$, medium for \textbf{$0.0588<\eta^2<0.1379$},  and large for \textbf{$\eta^2>0.1379$}. According to this criterion, $\eta^2$ values in  Table~\ref{tab:ANOVA_power} suggest that the difference in power between the two bursts is not only significant but also substantial.

\begin{figure*}
\baselineskip=12pt
\figline{\leftfig{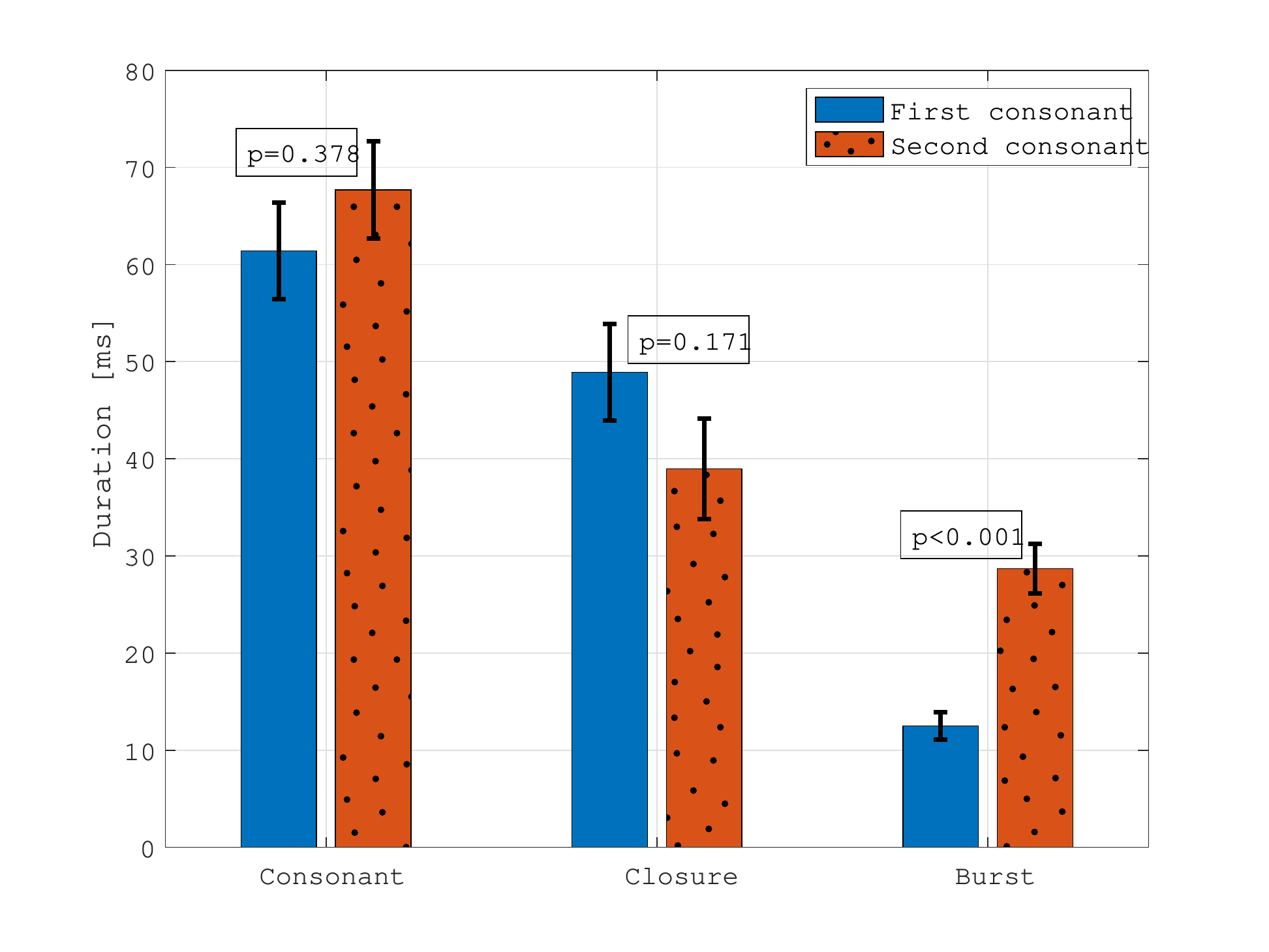}{0.5\textwidth}{(a) - Lexical gemination}\label{fig:Fig4a} \rightfig{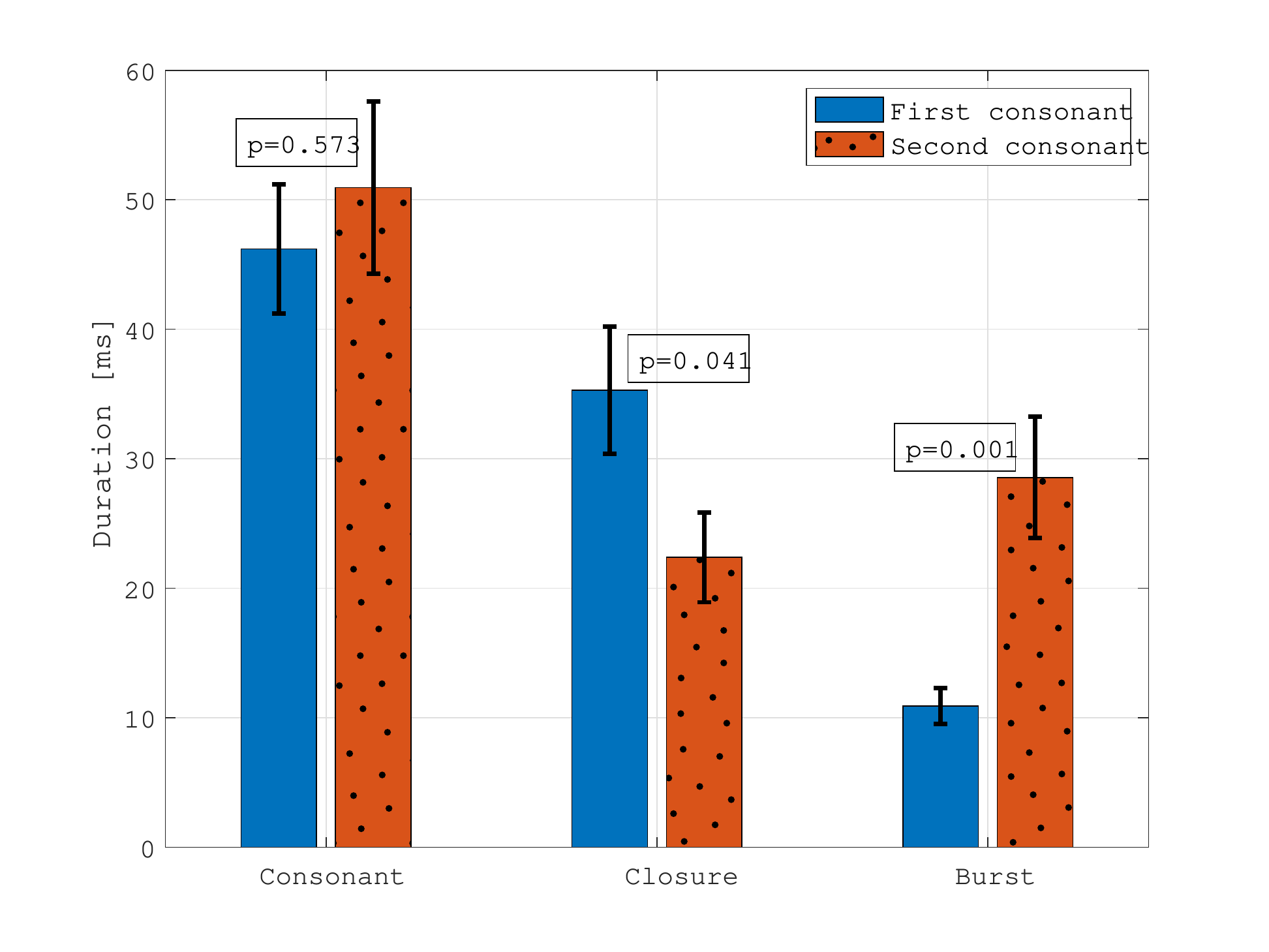}{0.5\textwidth}{(b) - Syntactic gemination}\label{fig:Fig4b}}
\caption{\label{fig:Fig4}Average value and standard error of durations of consonant, closure, and burst, of first (blue, plain) and second (red, dotted) consonants, in lexical geminates (Fig.~4(a)) and syntactic geminates (Fig.~4(b)). p-values on figure indicate significance level obtained with ANOVA tests. Bold values indicate significant difference, where the significance threshold was set at $p*=0.05$.}
\end{figure*}

\begin{figure*}[t]
\includegraphics[width=\textwidth]{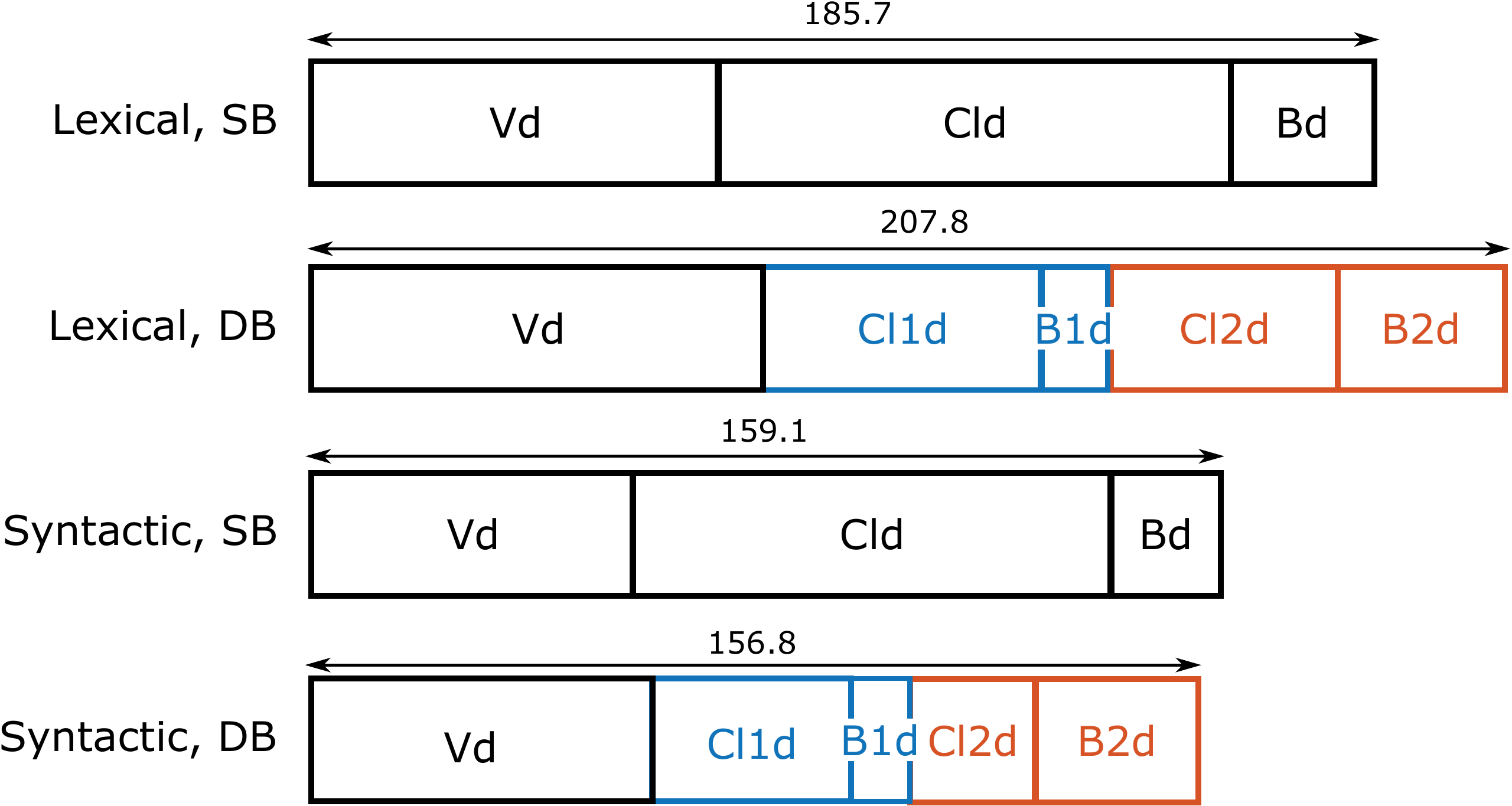}
\caption{\label{fig:Fig5}{Overall comparison of durations for consonant and preceding vowel in Lexical vs. Syntactic geminated stops, divided in Single Burst (SB) vs. Double Burst (DB). All values are expressed in ms. }}
\raggedright
\end{figure*}

In terms of durational parameters, the two consecutive consonants in both lexical and syntactic geminates with double bursts had similar duration. Although we believe that durational parameters refer to intervals between acoustic events rather than to duration of segments themselves, we will from now on refer, for the sake of simplicity, to durational parameters as segment durations. A univariate ANOVA test on consonant duration (duration of closure + duration of burst) with threshold $p*=0.05$  was carried out (see Table~\ref{tab:ANOVA_duration} in the Appendix for full results). Results highlighted a lack of statistical significance of this parameter: for lexical geminates $p=0.378>p*=0.05$ and for syntactic geminates $p=0.573>p*=0.05$. Conversely, burst durations were significantly different in the two consonants for both forms, with the $\mathrm{C^{(2)}}$ burst being much longer than the $\mathrm{C^{(1)}}$ burst: for lexical geminates $p<0.001<p*=0.05$ and a substantial difference ($\eta^2=0.345$), and for syntactic geminates $p=0.001<p*=0.05$ and a substantial difference ($\eta^2=0.318$), suggesting a time compensation between burst and closure, to keep consonant duration constant. A univariate ANOVA test on closure duration confirmed this prediction for syntactic geminates, in which closure duration of $\mathrm{C^{(1)}}$ was significantly higher than $\mathrm{C^{(2)}}$ ($p=0.041<p*=0.05$), but not for lexical geminates ($p=0.171>p*=0.05$), although also in this case we observed a systematic longer closure for the first consonant. The above results are summarized in Fig. \ref{fig:Fig4} (lexical in Fig.~4(a) and syntactic in Fig.~4(b)), that shows the average values and standard errors of consonant duration, closure duration, and burst duration.\\
As shown in the figure, average values were: a) for lexical geminates, $\mathrm{C^{(1)}}$ duration = 61.41 ms, $\mathrm{C^{(1)}}$ closure duration = 48.91 ms, $\mathrm{C^{(1)}}$ burst duration = 12.49 ms,  $\mathrm{C^{(2)}}$ duration = 67.66 ms, $\mathrm{C^{(2)}}$ closure duration = 38.97 ms, $\mathrm{C^{(2)}}$ burst duration = 28.7 ms; b) for syntactic geminates, $\mathrm{C^{(1)}}$ duration = 46.21 ms, $\mathrm{C^{(1)}}$ closure duration = 35.29 ms, $\mathrm{C^{(1)}}$ burst duration = 10.92 ms,  $\mathrm{C^{(2)}}$ duration = 50.94 ms, $\mathrm{C^{(2)}}$ closure duration = 22.38 ms, $\mathrm{C^{(2)}}$ burst duration= 28.56 ms. Very similar results were obtained for each speaker individually.\\
In summary, evidence for the presence of two consecutive consonants $\mathrm{C^{(1)}}$ and $\mathrm{C^{(2)}}$ was found for both gemination forms. In both forms, $\mathrm{C^{(1)}}$ burst power and duration were significantly weaker (lower power and shorter duration) than in $\mathrm{C^{(2)}}$ bursts.  $\mathrm{C^{(1)}}$  and $\mathrm{C^{(2)}}$ durations were similar for both forms. Closure duration was significantly longer in $\mathrm{C^{(1)}}$ than in $\mathrm{C^{(2)}}$ in syntactic geminates, but not in lexical geminates, although the observed tendency followed that same trend. The two consonants thus seemed similar overall in terms of duration but inter-event timing and power measurements support the hypothesis that the second consonant is stronger than the first.\\
A related question of interest is whether double burst consonants behave similarly to single burst consonants, in terms of durational parameters. A first parameter of comparison was burst duration. A different burst duration was observed above for $\mathrm{C^{(1)}}$ vs. $\mathrm{C^{(2)}}$, but the additional question was whether, on average, burst duration was similar in double vs. single burst consonants. The analysis was therefore based on computing the average of the duration of the bursts of $\mathrm{C^{(1)}}$  and $\mathrm{C^{(2)}}$ and comparing this average to the duration of the single burst in single burst consonants. Results of a univariate ANOVA test, with fixed factor being single burst vs. double burst consonant, showed no significant difference between the two groups ($p=0.160>p*=0.05$); see Table~\ref{tab:ANOVA_duration_burst} in the Appendix for complete results. Thus, the average duration of the two bursts of $\mathrm{C^{(1)}}$ and $\mathrm{C^{(2)}}$ was similar to the duration of the single burst in a single burst consonant. The average burst duration for double burst consonants was 20.3 ms vs. 23.3 ms for single burst consonants. In terms of consonant duration and closure duration, where closure duration for double burst consonants was computed as the sum of the two closures, two univariate ANOVA tests on these two parameters with fixed factor single vs. double burst consonant indicated a significant difference between the two groups for both consonant duration $p=0.033<p*=0.05$) and closure duration  ($p=0.026<p*=0.05$), although values were relatively close to threshold, and a small effect size was observed in the two tests (see Table~\ref{tab:ANOVA_duration_burst}). Average durations for single burst consonants were: consonant duration = 110.04 ms and closure duration = 86.71 ms, while for double burst consonants: consonant duration = 118.43 ms and closure duration = 77.81 ms. We further investigated the difference between single burst and double burst geminates by considering lexical and syntactic geminates separately. The observation on burst duration holds for syntactic gemination ($p=0.977, \eta^2<0.001$). For lexical gemination, the average duration of the two bursts was slightly dissimilar to the duration of a single burst in a single burst consonant ($p=0.044$), but the effect size was small ($\eta^2=0.017$), suggesting that the effect is minor. The same is not true for consonant duration and closure duration. In particular, consonant duration increased significantly in lexical geminates when a double burst was present, but this was not the case in syntactic geminates. In terms of closure duration, the opposite was observed; closure duration was not significantly changed by the presence of a double burst in lexical geminates, while it decreased significantly in syntactic geminates when a double burst occurred. The univariate ANOVA test for consonant duration, with fixed factor single vs. double burst consonant, provided the following quantitative data: a) for lexical gemination $p=0.001<p*=0.05$; b) for syntactic gemination $p=0.413>p*=0.05$. Average durations were: a) for lexical single burst geminated consonants = 114.77 ms and double burst geminated consonants = 129.07 ms; b) for syntactic single burst geminated consonants = 102.8 ms and double burst geminated consonants = 97.15 ms.  ANOVA univariate tests for closure duration with fixed factor single vs. double burst consonant provided the following quantitative data: a) for lexical gemination 	$p=0.81>p*=0.05$; b) for syntactic gemination $p<0.001<p*=0.05$ (see Table~\ref{tab:ANOVA_duration_burst} in Appendix for complete results). Average durations were: a) for lexical single burst geminated consonants = 89.09 ms and double burst geminated consonants = 87.88 ms; b) for syntactic single burst geminated consonants = 83.04 ms and double burst geminated consonants = 57.67 ms. To summarize, the average duration values are reported in Table \ref{tab:gemination_duration}.
\begin{table*}[t]
\caption{\label{tab:gemination_duration}Average duration in ms of time parameters for geminate stop consonants, divided by gemination type and single vs. double burst.}
\begin{ruledtabular}
\centering\footnotesize\setstretch{0.7}
\begin{tabular}{C{1.7cm}|C{1.6cm}|C{1cm}|C{1cm}|C{1cm}|C{1cm}|C{1cm}|C{1cm}|C{1cm}|C{1cm}|C{1cm}|C{1cm}}
 \multicolumn{2}{c|}{Gemination} & Vd & Cd & C1d & C2d & Cld & Cl1d & Cl2d & Bd & B1d & B2d\\
 \hline
 \multirow{3}{*}{Lexical} & SB & 70.9 & 114.8 & - & - & 89.1 & - & - & 25.7 & - & -\\
 	& DB & 78.8 & 129.1 & 61.4 & 67.7 & 87.9 & 48.9 & 39.0 & 41.2 & 12.5 & 28.7\\
	& Combined & 71.9 & 116.6 & - & - & 88.9 & -	 & - & 27.6 & 	- & -\\
	\hline
 \multirow{3}{*}{Syntactic} & SB & 56.3 & 	102.8 & - & - & 83.0 & - & - & 19.8 & - & -\\
 & DB & 59.7 & 97.2 & 46.2 & 50.9 & 57.7 & 35.3 & 22.4 & 39.5 & 10.9 & 28.6\\
 & Combined & 56.6 & 102.2 & - & - & 80.5 & - & - & 21.7 & - & -\\
\end{tabular}
\end{ruledtabular}
\end{table*}
In summary, single and double burst consonants were not substantially different in terms of neither burst duration nor consonant and closure durations. When lexical and syntactic geminates were considered separately, however, the analysis showed a different timing organization. In particular, in lexical geminates, closure duration was stable, but consonant duration increased for double burst consonants, i.e., the addition of a second burst had the effect of increasing consonant duration, since closure duration was stable. In syntactic geminates, the organizational pattern was different; consonant duration was stable, while closure duration was significantly decreased in double burst instances, suggesting that the second burst was inserted at the expense of closure duration. Figure \ref{fig:Fig5} summarizes the above comments and presents an overall comparison of the durations of consonant and its preceding vowel, highlighting the difference between lexical and syntactic gemination organizational pattern.\\
The above analysis highlighted the possibility of a different time planning  in lexical vs. syntactic gemination, when accommodating for a second burst. But was this the only behavioral difference between these two gemination forms? To come back to the initial question, were lexical geminates also acoustically different from syntactic geminates in other respects? \\
As mentioned in the Introduction, pre-consonant vowel duration has been shown in previous studies to be a relevant parameter in the acoustic manifestation of gemination of Italian consonants in general and in stops in particular \cite{EspDiB99}. Those studies showed that pre-consonant vowel duration was shortened while consonant duration was lengthened in geminate vs. singleton instances, and moreover that the ratio between consonant and pre-consonant vowel durations was a good predictor of the presence of gemination; geminated stops were typically characterized by a ratio of about 1.64 in VCCV words \cite{DiBDeN20b}. The ratio was usually much smaller for singleton consonants; in singleton VCV stops it was about 0.62 \cite{DiBDeN20b}. In the present study, the analysis of the ratio between consonant duration and vowel duration was extended by considering syntactic gemination, that can only be present in running speech, and by evaluating it both in combination with, and in comparison to, lexical gemination. To this aim, durational parameters and consonant duration to vowel duration ratio were also measured for the singleton stops of the LaMIT database; results are presented in Table \ref{tab:singleton_vs_geminate_duration}, together with data for geminates averaged over gemination type.\\
\begin{table*}[t]
\caption{\label{tab:singleton_vs_geminate_duration}Average values of durational parameters for singleton vs. geminate stop consonants of the LaMIT database, for speakers FS and MS. Durations are expressed in ms.}
\begin{ruledtabular}
\centering\footnotesize\setstretch{0.7}
\begin{tabular}{C{3cm}|C{2cm}|C{2cm}|C{2cm}|C{2cm}|C{2cm}}
  & Vd & Cd & Cld & Bd &  Cd/Vd\\
 \hline
 Singleton & 85.07 & 55.48 & 35.9 & 19.58 &  0.75\\
 Geminate & 65.98 & 111.01 & 85.69 & 25.32 & 1.84\\
\end{tabular}
\end{ruledtabular}
\end{table*}
The comparison between the ratios of single and geminate stops confirms the results for VCV vs. VCCV words reported in previous studies. The presence of gemination leads to a shorter vowel and a longer consonant, and thus to a ratio that increases from 0.75 in singletons to 1.84 in geminates; a threshold set to about 1 for the ratio, as proposed in \cite{DiBDeN20b}, is confirmed as a reliable discriminant for the presence of gemination. Moving to the comparison of ratios between gemination types, 
univariate ANOVA tests with lexical vs. syntactic gemination as a fixed factor indicated a significantly different ratio, with $p=0.004<p*=0.05$ (see Table~\ref{tab:ANOVA_ratios} in the Appendix for complete results) with a lower ratio value for lexical (average ratio = 1.76) than for syntactic (average ratio = 1.96), 
highlighting an additional difference in the acoustic manifestation of lexical vs. syntactic gemination. Results of statistical analyses also showed that within each gemination form, the ratio was stable when considering single vs. double burst consonants. In lexical gemination, the univariate ANOVA test, with fixed variable single vs. double burst, indicated a non-significant difference in the ratio ($p=0.9>p*=0.05$). The average ratio was 1.76 for single burst consonants and 1.78 for double burst consonants. A similar result was obtained in syntactic gemination with the following values: $p=0.216>p*=0.05$, average ratio for single burst consonants = 1.99, and average ratio for double burst consonants = 1.73 (see Table~\ref{tab:ANOVA_ratios_class} in the Appendix). The ratio between pre-consonant vowel and consonant durations proved therefore to be a crucial indicator of gemination in both gemination forms, with a higher ratio observed in syntactic gemination.

\section{Discussion of results and conclusions}
\label{sec:discussion}

Results of acoustic analyses showed that the acoustic manifestation of geminate consonants includes the presence, in some instances, of two bursts.  This is consistent with the hypothesis that the phonological representation of the geminate contains two consonants, whether two bursts are seen or not. This was observed in both lexical and syntactic gemination. 
This finding, and in particular the evidence for the presence of two bursts, provides strong support for a biphonematic nature of Italian geminated stop consonants and an answer to our first question: the acoustic manifestation of gemination in running speech was found to be not only related to durational parameters but was complemented by the discovery, in the acoustic signal, of double bursts and double closures, explicitly signaling the presence of a geminate, i.e. we may say a double consonant.\\
Furthermore, it was shown that the two consonants $\mathrm{C^{(1)}}$ and $\mathrm{C^{(2)}}$ did not typically differ in total duration but in the power and duration of the burst, as well as in closure duration, with a compensation effect between closure and burst durations. Since the first consonant $\mathrm{C^{(1)}}$ is characterized by a weaker burst (weaker power and shorter duration) than the second $\mathrm{C^{(2)}}$, it is plausible to say that the first consonant is less strong than the second. This observation supports the hypothesis that $\mathrm{C^{(1)}}$ is a coda consonant and $\mathrm{C^{(2)}}$ is an onset consonant, and therefore that $\mathrm{C^{(1)}C^{(2)}}$ form an heterosyllabic sequence. This finding answers our third question, on cluster syllabification. This observation also provides additional evidence that relates to the planning of timing of the production process. In syntactic gemination, the presence of the second burst only impacted closure duration; it did not influence consonant duration. That is, the extra burst was accommodated in the closure time interval. But this was not the case in lexical gemination, for which closure duration was kept stable when the extra burst was included by simply making the consonant longer. This finding answered our second question, and indicated that the two gemination forms, lexical vs. syntactic, may arise at two different points during the production planning process. In syntactic gemination in particular the phenomenon must arise after words have already been planned, since the phenomenon occurs across words; this may explain why syntactic gemination may not alter the duration of the onset consonant. In contrast, lexical gemination happens within a word, and, as such, timing elements in the word may still find room for adjustments.\\
Finally, the ratio between consonant and pre-consonant vowel durations was analyzed since this ratio was shown in previous studies focusing on VCV vs. VCCV words to be a good indicator for the presence of gemination. This result was confirmed for the running speech material provided in the LaMIT database: a ratio above 1 typically indicates the presence of gemination.\\
In the present study, results showed that the above ratio was stable across single burst vs. double burst groups, for both lexical and syntactic geminates, with ratio values higher than in previous studies on VCV-VCCV words, i.e. in the direction of reinforced gemination, manifested by multiple cues that may reveal the presence of the second consonant. This concept may recall that of enhancement and leads to a model by which the geminated consonant is made of two consonants and appears in the acoustic signal as a longer consonant (because they are two consonants), an additional durational cue being the shortening of the pre-consonant vowel (because the signal containing that vowel is a closed syllable) and the presence of the burst. This finding indicates that the ratio is a stable parameter and that the insertion of a second burst does not alter the rhythmic structure of a word (lexical gemination) or the rhythmic structure across words (syntactic gemination). The finding that an additional burst is not always visible in geminated consonants paves the way to an interesting research question: is a missing extra burst the result of occasional articulatory failure in introducing an extra cue reflecting the intention of the speaker? The fact that an extra burst is not always visible in the repetitions of an identical word seems to support this interpretation. It will be interesting to also test this hypothesis by investigating the relation between the  occurrence of a second burst and speaking rate. On the other hand, some stops seem to be produced with this additional cue more often than others when geminated (/k/ vs. /d/). Future research will also focus on this aspect, together with gathering articulatory data. In a previous instrumental investigation Lehiste et al. (\cite{Leh73}) found electromyographic evidence for rearticulation in both intervocalic geminate consonants and junctural C+C sequences in Estonian, a language that has opposition between single and geminate consonants in intervocalic position. In the same study, Lehiste et al. found no evidence for rearticulation in English junctural C+C sequences, which suggests that the articulation of gemination may follow language-specific patterns. In a cinematic study of bilabial and labiodental articulation in Italian, using optoelectronic measurements, Zmarich and Gili-Fivela (\cite{ZmaGil05}) observed that despite ambiguous acoustic manifestations, cinematic correlates of gemination were similar to correlates of heterosyllabic consonant clusters, further supporting the biphonematic status of geminates in Italian.\\
The analysis also showed an additional difference between the two gemination forms: the ratio was significantly different between lexical and syntactic gemination, and, in particular, it was higher for syntactic gemination, mainly due to a shorter pre-consonant vowel duration. \\
The above difference may be due to the different nature of the two geminations, lexical gemination being always contrastive while syntactic gemination is almost always not. Is the expressive nature of syntactic gemination the reason for a shorter pre-consonant vowel leading to a reinforced ratio? Or are there other driving factors to justify vowel shortening? The interpretation of  this finding may require additional experiments focused on this specific aspect, possibly addressing those rare but existing cases where, in Italian, syntactic gemination becomes contrastive. \\
Beyond addressing the natural extension of the analysis of lexical vs. syntactic gemination in other consonant classes, future work will further focus on the effect of the proposed heterosyllabic structure /$\mathrm{C^{(1)}.C^{(2)}}$/ on coarticulation, and analyze in particular whether the extent of coarticulation between different speech segments is different across geminate vs. singleton consonants, as suggested by previous studies that address the effect of consonant cluster syllabification on vowel-to-vowel coarticulation \cite{Mok12}. \\
As a final remark, our results were based on only two speakers. However, our interpretation is supported by the fact that a similar ratio of double bursts was found for the two speakers, and, furthermore, in a few cases, an identical word by an individual speaker, in the two repetitions, showed two bursts in one case and a single burst in the other. In these two types of tokens, the duration of the preceding vowel is similar, suggesting that speakers may have similar phonetic and phonological representation of geminate consonants, whether two bursts are visible or not. Testing this conclusively will require, however, further experimentation on a larger set of speakers.\\

\begin{acknowledgments}

This work was supported in part by the Radcliffe Institute for Advanced Study at Harvard University and by Sapienza University of Rome within the research project “Towards Speech Recognition of the Italian Language Based on Detection of Landmarks and Other Acoustic Cues to Features”, grant \# RP11916B88F1A517 and RP120172B3612D94.

Stefanie Shattuck-Hufnagel gratefully acknowledges the support of the National Science Foundation, grant \# BCS 1827598.

\end{acknowledgments}



\appendix*
\section{Statistical analysis data}

\begin{table*}[t]
\caption{\label{tab:ANOVA_power} Degrees of freedom, test variable F, probability p at which the null hypothesis can be rejected and effect size estimation $\eta^2$ obtained in the univariate ANOVA tests performed on power of first vs. second burst in words containing double bursts. The fixed factor is first burst vs. second burst. Three tests were run for: lexical gemination, syntactic gemination, and lexical and syntactic combined; bold characters indicate significantly different values, with threshold set as p*=0.05.}

\begin{ruledtabular}
\begin{tabular}{ccccc}
 Type & Degrees of freedom& F &p& $\eta^2$ \\
\hline
Lexical & (1, 59) & 17.056 & \bf{$<$0.001} &0.227\\
Syntactic & (1, 29) & 8.227 & \bf{0.008} &0.221\\
Combined & (1, 89) & 25.012 & \bf{$<$0.001}&0.221 \\
\end{tabular}
\end{ruledtabular}
\end{table*}

\begin{table*}[t]
\caption{\label{tab:ANOVA_duration}Degrees of freedom, test variable F, probability p at which the null hypothesis can be rejected and effect size estimation $\eta^2$ obtained in the univariate ANOVA tests performed on durational parameters of first consonant $\mathrm{C^{(1)}}$ vs. second consonant $\mathrm{C^{(2)}}$ in words containing double bursts. The fixed factor is first burst vs. second burst. Two tests were run for lexical gemination and syntactic gemination; bold characters indicate significantly different values, with threshold set as p*=0.05.}

\begin{ruledtabular}
\begin{tabular}{cccccc}
Type & Parameter& Degrees of freedom & F &p  & $\eta^2$\\
\hline
\multirow{3}{*}{Lexical} & Consonant duration & (1, 59) & 0.788 & 0.378& 0.013\\
& Closure duration & (1, 59) & 1.920 & 0.171 &0.032\\
& Burst duration & (1, 59) & 30.562 & \bf{$<$0.001}& 0.345\\
\hline
\multirow{3}{*}{Syntactic} & Consonant duration & (1, 29) & 0.325 & 0.573 & 0.011\\
 & Closure duration & (1, 29) & 4.602 & \bf{0.041}& 0.014\\
& Burst duration & (1, 29) & 13.064 & \bf{0.001} &0.318\\

\end{tabular}
\end{ruledtabular}
\end{table*}

%
%

\begin{table*}[t]
\caption{\label{tab:ANOVA_duration_burst}Degrees of freedom, test variable F, probability p at which the null hypothesis can be rejected and effect size estimation $\eta^2$ obtained in the univariate ANOVA tests performed on consonant duration Cd, closure duration Cld and burst duration Bd for single burst words vs. average burst duration for double burst words. The fixed factor is presence of a single burst vs. a double burst. Three tests were run for: lexical gemination, syntactic gemination, and lexical and syntactic combined; bold characters indicate significantly different values, with threshold set as p*=0.05.}

\begin{ruledtabular}
\begin{tabular}{cccccc}
Type & Parameter& Degrees of freedom & F &p  & $\eta^2$\\
\hline
\multirow{3}{*}{Lexical} & Cd & (1, 239) & 10.545 & \bf{0.001}& 0.042\\
& Cld & (1, 239) & 0.058 & 0.810 &$<$0.001\\
& Bd vs. average Bd  & (1, 239) & 4.103 & \bf{0.044} & 0.017\\
\hline
\multirow{3}{*}{Syntactic} & Cd & (1, 151) & 0.675 & 0.413 & 0.004\\
 & Cld & (1, 151) & 17.492 & \bf{$<$0.001}& 0.104\\
 & Bd vs. average Bd  & (1, 151) & 0.001 & 0.977 &$<$0.001\\
\hline
\multirow{3}{*}{Combined} & Cd & (1, 391) & 4.575 & \bf{0.033} & 0.012\\
 & Cld & (1, 391) & 5.018 & \bf{0.026}& 0.013\\
 & Bd vs. average Bd  & (1, 391) & 1.986 & 0.160 &0.005\\
\end{tabular}
\end{ruledtabular}
\end{table*}

\begin{table*}[t]
\caption{\label{tab:ANOVA_ratios}Degrees of freedom, test variable F, probability p at which the null hypothesis can be rejected and effect size estimation $\eta^2$ obtained in the univariate ANOVA tests performed on previous vowel duration Vd, consonant duration Cd and on the ratio Cd/Vd. The fixed factor is the gemination type (lexical vs. syntactic); bold characters indicate significantly different values, with threshold set as p*=0.05.}

\begin{ruledtabular}
\begin{tabular}{ccccc}
Parameter & Degrees of freedom  & F &p & $\eta^2$\\
Vd &(1, 391) & 47.997 & \bf{$<$0.001}& 0.11\\
Cd &(1, 391) & 33.398 & \bf{$<$0.001}& 0.079\\
Cd/Vd &(1, 391) & 8.521 & \bf{0.004}& 0.021\\

\end{tabular}
\end{ruledtabular}
\end{table*}

\begin{table*}[t]
\caption{\label{tab:ANOVA_ratios_class}Degrees of freedom, test variable F, probability p at which the null hypothesis can be rejected and effect size estimation obtained in the univariate ANOVA tests performed on the ratio between consonant duration Cd and previous vowel duration Vd. The fixed factor is presence of a single burst vs. a double burst. Two tests were run for lexical gemination and syntactic gemination; bold characters indicate significantly different values, with threshold set as p*=0.05.}

\begin{ruledtabular}
\begin{tabular}{cccccc}
Type & Degrees of freedom & F &p & $\eta^2$\\
 Lexical & (1, 239) & 0.015 & 0.904& $<$0.001\\
 Syntactic & (1, 151) & 1.547 & 0.216& 0.01\\

\end{tabular}
\end{ruledtabular}
\end{table*}
\clearpage
\bibliography{Manuscript.bib}


\end{document}